\documentclass[a4paper,10pt]{article}

\usepackage{graphicx,setspace}
\usepackage{bm}
\usepackage{amsmath,textcomp}

\begin{document}

\title{Fragmentation, integration and macroprudential surveillance of the US financial industry: Insights from network science}

\author{Y\'{e}rali Gandica$^{1,2,\ast}$, Marco Valerio Geraci$^{1,3}$, Sophie B\'ereau$^{1,4}$,Jean-Yves Gnabo$^{1,2}$ \\
\emph{1. CeReFiM, Universit\' e de Namur, Namur, Belgium. } \\
\emph{2. Namur Center for Complex Systems - naXys, Universit\' e de Namur, Namur, Belgium.} \\ 
\emph{3. ECARES, Universit\'{e} libre de Bruxelles, Bruxelles, Belgium.} \\
\emph{4. CORE, Universit\'{e} Catholique de Louvain, Louvain-la-Neuve, Belgium.} \\
}
\footnotetext{$^{1}$Correspondence author. E-mail: ygandica@gmail.com}
\date{\today}

\maketitle
\singlespacing
\begin{abstract}

Drawing on recent contributions inferring financial interconnectedness from market data, our paper provides new insights on 
the evolution of the US financial industry over a long period of time by using several tools coming from network science. 
Following \cite{Geraci} a Time-Varying Parameter Vector AutoRegressive (TVP-VAR) approach on stock market returns to retrieve unobserved 
directed links among financial institutions, we reconstruct a fully dynamic network in the sense that connections are let to 
evolve through time. The financial system analysed consists of a large set of $155$ financial institutions that are all the banks,
broker-dealers, insurance and real estate companies listed in the Standard \& Poor's $500$ index over the period $1993 - 2014$. Looking
alternatively at the individual, then sector-, community- and system-wide levels, we show that network science’s tools are able 
to support well known features of the financial markets such as the dramatic fall of connectivity following Lehman Brothers’ 
collapse. More importantly, by means of less traditional metrics, such as sectoral interface or measurements based on contagion 
processes, our results document the co-existence of both fragmentation and integration phases between firms independently from the
sector they belong to, and in doing so, question the relevance of existing macroprudential surveillance frameworks which have been
mostly developed on a sectoral basis.
Overall, our results improve our understanding of the US financial landscape and may have important implications for risk monitoring
as well as macroprudential policy design.
\end{abstract}

\footnotetext{$^{1}$Correspondence author. E-mail: ygandica@gmail.com}
\newpage \baselineskip1.0cm
\singlespacing

\section{Introduction}
The strong interdependence among financial institutions has been emphasised in numerous academic contributions, mainly after 
the $2007-2008$ worldwide financial crisis, leading to a gradual shift from a micro to a macroprudential approach to financial 
stability.\footnote{According to the Financial Stability Board (FSB) and the Basel Committee for Banking Supervision (BCBS), 
financial interconnectedness is a determinant of systemic risk along with cross-jurisdictional activity, size, substitutability 
and financial institution infrastructure and complexity. It is defined as the network of contractual obligations which can 
potentially channel financial distress \cite{Basel}} Yet, the progress made to exploit the potential
of this  system-wide framework is still limited compared to other fields, such as physics, neurology or biology to quote only
a few in which network and complex systems' representations have been deeply rooted for decades  
\cite{bouchaud2008,farmer2009}.\footnote{As discussed in \cite{bouchaud2008,farmer2009} one of the fruitful consequences of the 
financial crisis was the scientific opening to understand and accept the view of an economic and a financial landscape as a 
complex system. However, this approach is still in its infancy compared to other disciplines in which specific tools have been
developed to analyse complexity.} Given the importance of financial stability  on  economic growth and welfare,
it seems critical to pursue the effort in this direction and to accelerate the \textquotedblleft transfer of 
knowledge\textquotedblright\ from other fields to expend our understanding of financial systems.  
 
Against this background, the aim of this paper is twofold. On the one hand, it brings up-to-date knowledge from network 
science to economic and financial systems. More specifically, we use a large set of tools devoted to the analysis of network'
structures in order to  improve our understanding of the US financial industry over a long time span. By doing so, we are able
to identify  phases of integration and fragmentation in the industry, both within as well as across financial sectors. As
such, our results can document among others the   debate regarding the relevance of sector-based  metrics and supervision
authorities for macroprudential surveillance.\footnote{In 2011, the Bank of New York Mellon, State Street Corporation, and
Northern Trust Corporation (collectively “Specialized Custody Banks”) addressed the following concern to the Basel Committee 
on Banking Supervision: “In its assessment of interconnectedness category as used [for measuring systemic importance of banks],
we recommend that the Proposal better define ‘Financial Institution’ in designating the assets and liabilities to be included 
in these indicators. For example, it is unclear the extent to which the definition of ‘Financial Institution’ includes 
collective investment vehicles such as mutual funds, collateral investment pools, or other types of private funds and
investment vehicles.”} On the other hand, the specificities of our dataset, as detailed below, and in particular the 
observation of a highly dynamic system over a long period of time, are intended to provide new insights to network science. 
 
One of the greatest difficulties faced by researchers when analysing financial networks lies in the lack of reliable and 
comprehensive datasets on the “physical” relationships between institutions. To deal with this issue, a now common approach in
the literature consists of making use of information embedded in market-based data as done in \cite{Billio,Diebold}. In a
nutshell, financial institutions represent the nodes of the network with the linkages reflecting the relative influences
between pairs of firms, which corresponds for instance to how much stress for an institution as materialised by severe losses, 
is transmitted to another institution. The identification of those links is based on the statistical measurement (be it 
correlation or other form of causal measure, for instance) of temporal dependences between one or several observable 
characteristics associated to the nodes, such as their stock market returns. Relying on this approach to form pairwise 
connections instead of observed physical contracts or transactions between institutions implies that the information 
propagating over the system, far from being innocuous, strongly affects all the nodes in the path. In other words, the links 
of our system reflect “effective” transmission channels of financial distress (i.e. based on changes in the firm’s asset 
value) as opposed to “potential” channels stemming from physical connections, such as cross-lending or common portfolio 
holdings. In this paper, we follow a market-based approach to retrieve the network representation of the financial system 
whose characteristics will then be studied by means of several network metrics successively considering individual, then 
sector-, community- and finally system-wide analyses. More specifically, we apply the “causal” version of this approach in 
which statistical dependences are assessed with time lags to recover the directionality of the relationship and in turn, 
separate influencer from receiver nodes. Another important aspect of our system is that nodes are institutions belonging to 
four different financial sectors, i.e. banks, broker-dealers, insurance companies and real estate companies. As a consequence,
we are able to study whether the level of potential disruptions caused by any propagation mechanisms may depend, not only on 
the structural position of the nodes in the network (i.e. their level of criticality), but also on the sectors those nodes 
belong to.  Importantly, our network is fully dynamic in the sense that it is analysed over
successive periods with potential changes within its structure over time. Last but not least, we go beyond sector-based categories to identify sub-groups of institutions by using a more agnostic approach consisting in applying a community detection algorithm which allows to recover \textquotedblleft data-driven\textquotedblright\  groups of highly connected financial institutions.  

Following \cite{Geraci}, the linkages  between financial institutions  are retrieved from stock market prices by means of
Bayesian Time-Varying VAR framework in the vein of the one developed by \cite{Primiceri} for macroeconomic data.  One of the 
main advantages of this methodology is that it allows to generate temporal networks in a widely spaced period of $20$ years, 
where connections are evolving gradually through time. Another powerful characteristic is that it does not use the common
rolling window approach, where causal relationships are estimated over successive sub-samples which may lead to several 
empirical problems that may blur the identification of real linkages between nodes. Instead, statistical inference is
established by taking into account the whole temporal spectrum of the data, therefore overcoming some limitations of 
rolling-window approaches, approaches, regarding sensitivity to window sizes and outliers. This methodology is applied to a
large dataset that can reasonably be considered as an adequate representation of the US financial system over two decades as 
it embeds all the financial firms listed in the S\&P 500 index, belonging to the four previously mentioned sectors. Overall, 
it includes $155$ financial institutions for which we observe monthly stock prices from April $1993$ to November $2014$. Working at a 
relatively low frequency allows avoidance of potential problems due to stock market noise and eases the identification of
causal linkages.  This analysis is performed at the system wide, sectoral and community levels. For the former approach, we apply the so-called \textquotedblleft Louvain method\textquotedblright\ on our networks at each period considering undirected edges. By doing so, we recover a set of communities within which financial  institutions could influence each others. The \textquotedblleft Louvain method\textquotedblright\    has been widely used in the network science literature to isolate groupes of highly connected nodes.

Equipped with this dynamic representation of the US financial industry, our main goal is then to
describe its characteristics with a set of metrics taken from network science. Accordingly, we do not
intend here to provide an in-depth analysis of the impact of network topology on market stability, nor do we document the 
driving factors of network formation. Such avenues in a context in which the dynamic nature of financial systems has been 
properly featured are left for future research. We believe, however, that this contribution provides important insights to 
complement existing studies that use network representations to analyse the financial industry. Among the most related 
contributions, we can quote \cite{Billio,Diebold}. In their pioneering works, they propose a way to overcome the absence of 
comprehensive data on “physical” relationships between financial institutions, by inferring the network from stock market data.
Applying the Granger-causality test on bivariate models \cite{Billio} or more sophisticated Vector 
Autoregressive model and variance decomposition \cite{Diebold}, they provided the first set of empirical evidence regarding 
the structure of interconnectedness for the financial industry, including banks, insurance companies or investment funds. One
of the main caveats of this generation of studies lies in the way the temporal nature of financial networks was dealt with as
they did rely on successive snapshots of the industry at different time periods by the mean of rolling windows regressions. 
Recently, there have been a few studies among which \cite{Geraci,Billio2} some have developed methodological approaches 
specifically designed to better tackle the time varying dimension of the financial system and infer fully dynamic networks 
\cite{Geraci,Billio2}. However, these contributions do not exploit in full the set of tools offered by network science to 
document salient patterns, relying on degree centrality measures. We propose to fill the existing gap in this contribution by
addressing these two caveats together.

By doing so, we more specifically contribute to the literature by  shedding a new light on  the following issues : (i) the 
influence/vulnerability of financial institutions at the individual  and at the sector  level, and (ii) the
fragmentation/integration of the whole financial industry. For this, we concentrate on six specific aspects of the network :
(i) degree centrality, (ii) community structure, (iii) component structure, (iv) sector interface, (v) katz centrality and 
M-reach centrality (contagion process), (vi) \textquotedblleft top\textquotedblright\  institution behavior.\footnote{
At each time period, we define as top institutions, those having a centrality measure greater than $85 \%$ of the highest	
value in that period, that is for which the value falls into a range defined by the highest value on one side and a threshold 
equal to $85\%$ of the highest value on the other side. The centrality is calculated by using alternative measures throughout
the study, such as in-degree, out-degree or betweenness to cite a few. Then, we count the number of institutions belonging to 
the “top” list for each sector. This measure is inspired by the one developed in [9]. It is simple to implement and enables
easy comparison of the level of connectedness of the different sectors when considering only the contribution of the most 
connected institutions, that we call “top” institutions. This local measure also reflects how much the most central 
institution of the system that is the most risky or the most vulnerable is isolated or if there is group of highly connected
institutions during the period at stake.}
 
Our results document  20 years of the evolution of US financial industry. The critical role played by the banking  and the 
real estate sectors in the successive episodes of financial turmoil appears clearly, the former emerging as a major 
transmitter of risk in the industry, while the latter appears as the main absorber. The insurance companies also appear as
central because of both their exposure to the rest of the system and their effect on other institutions. The role of 
broker-dealers conversely has been  more moderated than the three other industries all over the sample. Our results also 
provide important evidence regarding the co-existence of both fragmentation phases within the financial industry along with
overall increasing integration among firms, in whatever the sector they belong, which, as a result, questions the accuracy of 
sector-based macroprudential frameworks \footnote{Current supervision system is articulated around Basel III for banks, 
Solvency II for insurance companies and the Markets in Financial Instruments Directive 2004/39/EC (known as ``MiFID") for 
European investment funds for instance.}.
From a network science perspective, our results illustrate the relevance of the traditional tools developed in this field for
analysing “spillover-based” financial networks. Specifically, it shows that relevant information can be retrieved from 
financial data with centrality measures at the node- and sector-level. It provides evidence that specific group-detection 
approaches such as community structure or component structure detection algorithms are of interest to create “agnostic” 
categories of financial institutions in addition to sector-based groups. It documents the quasi equivalence on financial 
data of alternative measures such as contagion process and the Katz centrality \footnote{We set the number of steps for the 
contagion process (M-reach centrality) to two, and the damp parameter to 0.625 for the Katz centrality.}. Lastly, we confirm 
the interest of extracting information from top values of centrality measures as done among others in \cite{Gandica} in the context of 
Wikipedia page edition. 

The remainder of the paper is the following. We describe in the second section the construction of the network. The third 
section discusses the results. We start with an analysis of the network’s main characteristics at the sectoral, community and 
component levels. Then, we examine sectoral interface and eventually discuss tools related to contagion modelling. Finally,
the fourth section concludes.

\section{Methods}
To represent the interdependencies at stake in the financial system, we follow \cite{Geraci} who propose a 
framework based on on time-varying parameter vector autoregressions, as in \cite{Primiceri,Cogley} to recover a network of 
financial spillovers  - or causality-based network - that is entirely dynamic. In their framework, financial institutions 
represent nodes in a directed network. Whereas spillovers, measured as temporal dependence between the stock price returns of 
the financial institutions, represent the directed edges of the network.
Temporal dependence between stock returns is measured according to the following vector autoregression:
\begin{equation}
R_t =c_t +B_t R_{t-1} +u_t \ 
\label{eq:1}
 \end{equation} 
\noindent where $R_t = [r_{1t} ,...,r_{Nt}]^{'}$  is the vector of the stock returns of the $N$ financial institutions in the network. 
$c_t$ is the time-varying intercept, whereas $B_t$ is a $N$x$N$ matrix of time-varying autoregressive coefficients, which determines
the temporal dependence between the stock returns and therefore the directed spillovers between the financial institutions. 
\\

Precisely, a directional edge is drawn at period $t$ from $i$ to $j$, if the  $ji$ element of $B_t$,  $B^{(j i)}_t$, is 
significantly different from zero. The framework parallels the classic time-invariant approach of recovering financial 
spillover networks using Granger causality, see e.g., \cite{Billio}. 

Finally, the errors, $u_t$, are assumed to be normally distributed, with mean zero and variance-covariance matrix $\Sigma$
(see Appendix SA of the Supporting Information). The original model of  \cite{Geraci} allows for heteroskedasticity and 
fat-tailed errors. Here, however, we adopt a simpler approach and standardize (so to have unit variance) the returns, 
$r_{1t} ,...,r_{Nt}$, in a previous step by using a GARCH(1,1) model to estimate the time-varying volatility.

The model in Eq. \ref{eq:1} can be re-written in a compact-form, as
\begin{equation}
R_t = X_t^{'}\theta_t +u_t,
\label{eq:2}
 \end{equation} 

\noindent where $X_t= I_N \otimes [1,R_{t-1}]$ and $I_N$ is a $N$x$N$ identity matrix. $\otimes$  is the Kronecker product,
and $\theta_t$ is a vector with the stacked elements of $c_t$ and $B_t$.

The time-varying parameters are then assumed to evolve according to a random walk:
\begin{equation}
\theta_t= \theta_{t-1} + \nu_t,
\label{eq:3}
\end{equation} 

\noindent where $\nu_t \sim N(0,Q)$.

This assumption allows the time-parameters, $c_t$ and $B_t$ to evolve flexibly over time and to allow the data to speak 
by itself. The amount of time-variation, is governed by the variance of the errors, $Q$, which is estimated along with 
the other parameters of the model, $\theta_t$ and $\Sigma$. 

Finally, in order to determine the existence of a link from $i$ to $j$, at a given time period t, we test the following null
hypothesis:

\begin{equation}
H_{0,t}:B^{(ji)}_{t}=0 \ \forall j \neq i \\ 
\label{eq:4}
\end{equation}

The model outlined in equations \ref{eq:1}, \ref{eq:2} and \ref{eq:3} is estimated using Bayesian techniques following
\cite{Calomiris}. Then, the hypothesis given by equation \ref{eq:3}, is tested using Bayesian inference. Specifically, we use 
Bayes Factor, which gives the odds in favor of the null hypothesis against the alternative hypothesis, 
$H_{1,t}:B^{(ji)}_{t} \ne 0$, without assuming that the null hypothesis is true. Bayes Factor is estimated following \cite{Koop}. Once we retrieve Bayes Factor, we look at the implied probability that  is
true \footnote{If $\hat{K}^{ji}_{t}$ is Bayes Factor for $H^{(ji)}_{0,t}$, then the implied probability is just 
$\hat{K}^{ji}_{t} / (1+\hat{K}^{ji}_{t})$}.

We use the implied probability to retrieve the network at different cut-off points. Effectively, the cut-off is a filtering 
mechanism and a higher cut-off leads to a more dense network with more links. The stability of our analysis is assessed by 
varying the cutoff levels of the statistical test used for detecting the links. In most of the figures throughout the study,
the four following cutoffs are considered to be: 5\%, 7\%, 10\% and 15\%. 

The prior distributions assumed for the parameters to retrieve the Bayesian estimates, are given in Appendix SA1. The 
posterior distribution algorithm, which, for the time-varying parameters, uses the Kalman filter and smoother as per Carter and Kohn 
(1997), is given in Appendix SA2. 

We applied the model to all financial institutions among banks, insurers and real estate companies (SEC codes 6000 to 6799) 
that were components of the S\&P 500 between January 1990 and December 2014. For these companies we collected the stock price 
at monthly close from Thomson Reuters Eikon over the same time period. Initially the sample contained 182 firms but was 
reduced to 155 after restricting our analysis to stocks with at least 36 monthly observations. As mentioned, all returns were 
standardized using a GARCH(1,1) model, to account for heteroskedasticity, prior to applying the time-varying framework 
highlighted above.
\\

\begin{figure}[!h]
\centering
\includegraphics[width = 13cm]{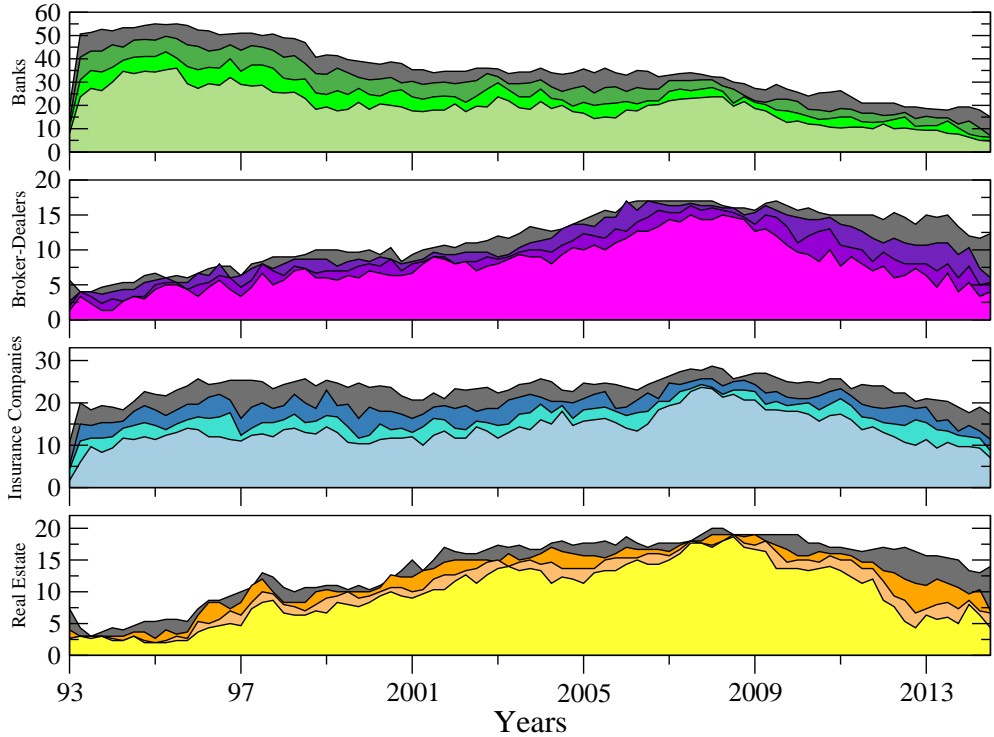}
\caption{{\bf Number of connected nodes per financial sector (i.e. sub-networks) from April 1993 to November 2014.} From top 
to bottom: banks (green), broker-dealers (purple), insurance companies (blue) and real estate companies (yellow). The different
tonalities in each plot are related to the sensibility parameter (i.e. test's cutoff levels  for detecting  significant links)
set to 5 \% (softest tonality), 7\%, 10\% and 15\% (dark grey) respectively. }
\label{fig:1}
\end{figure}

\section{Results}
\subsection{Sector-based behavior}
In this section, we provide an overview of the network. The number of nodes in the system as well as centrality measurements 
are displayed by sector.  
  
\begin{figure}[!h]
\centering
\includegraphics[width = 14cm]{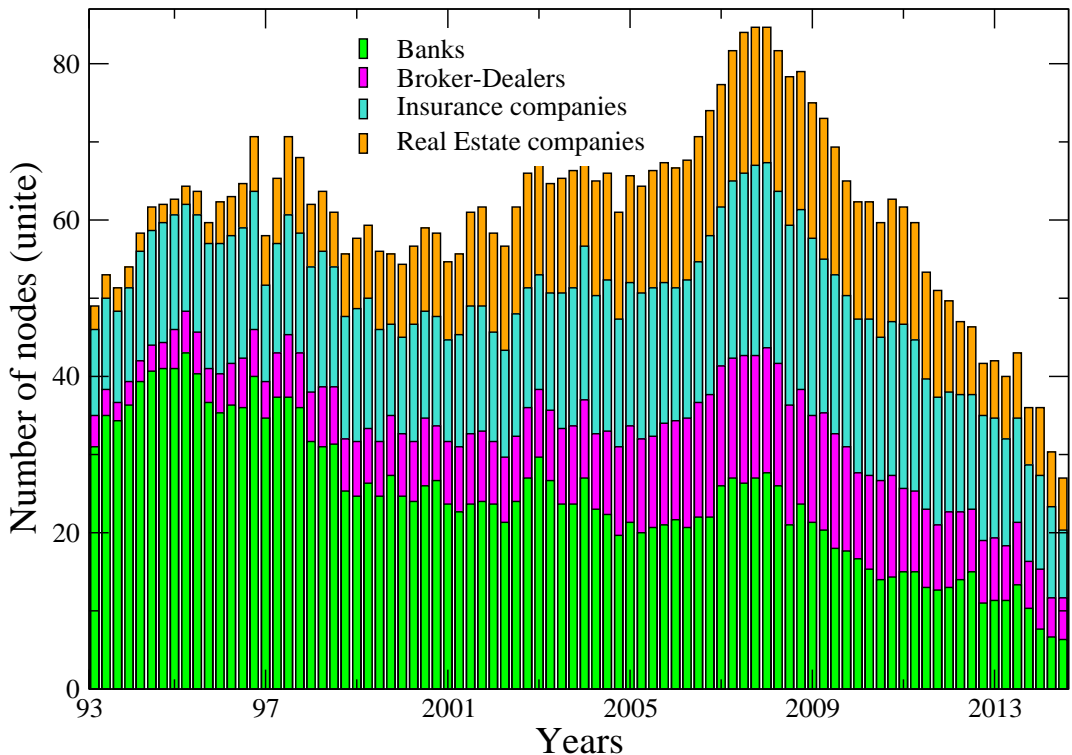}
\caption{{\bf Each bar size represents the number of connected institutions within our four sectors, from April 1993 to November 2014}: banks (green), 
broker-dealers (purple), insurance companies (blue) and real estate companies (yellow). Values are displayed at quarterly 
frequency to ease the visualisation. The sensibility parameter has been fixed to 10\%.} 
\label{fig:2}
\end{figure}

Fig.~\ref{fig:1} shows the number of connected nodes comprising our network across time among the different sectors mentioned 
above. We can notice, from a first visual inspection, that the main patterns characterising the evolution within each sector 
are not sensitive to the choice of the cutoff level used to detect the links. Comparing now the series across sectors,
Fig.~\ref{fig:1} exhibits contrasting dynamics. In particular, while the number of connected banks has steadily decreased since 1995, insurance companies and real estate companies experienced two successive phases characterised by an upward trend until late 2008 followed 
by a reversal. The number of broker-dealers overall displays more stability. It is worth recalling at this stage that as our 
connections are built on causal relationships among asset prices; an institution identified as not connected to rest of the 
network can to some extend be physically related to other institutions. However, such relationship does not materialise into 
spillover effects. Fig.~\ref{fig:2} completes the picture by adding up the number of nodes across sectors \footnote{For the sake of clarity, we will often show across the paper the results related to the sole $10 \% $ cutoff level. 
The figures for other significance cutoff levels are available upon request.}. We can more clearly observe from there the various phases at the system-wide level. 
From 1993 to 1997, the number of connected nodes sharply increased, being mainly driven by the banking sector. The trend reversed then until 2001, reaching 
a total of 55 connected nodes from 70 in 1997.  From 2001 to 2008, the upward trend resumes to attain the highest level of the
whole sample by peaking at more than 80 nodes. Interestingly, as noted in Fig.~\ref{fig:1}, the number of banks decreased over
that period. Hence, our data reveals the presence of a shift among the set of large connected institutions from the US 
financial industry in the run-up to the financial crisis with insurance companies as well as real estate companies, and to a
lower extend broker-dealers, gaining in importance within the system. Eventually, the size of the network dramatically shrunk 
after 2008, reaching its lowest level at around 35 connected nodes - in our case the number of institutions influencing or been influenced 
by the rest of the system - institutions in late 2013. The four sectors were almost equally affected by this drop. Such a result implies that many institutions became isolated or weakly connected to the rest of the system
\footnote{It is worth reminding that our identification strategy for the links is based on 
 statistical tests. Therefore, the absence of a link between pairs means that we do not have enough evidence, or, said 
 differently, the evidence is too weak
 in the data to reject that institutions could be independent.}.

 From this first inspection, we can notice that the dates of the successive turning points correspond to well-known financial 
 events such as the Asian crisis in late 1997, the burst of the so-called \textquotedblleft dot.com\textquotedblright\ bubble
 in 2001 as well as the  $2007$ - $2009$ financial crisis.  \\

\begin{figure}[!h]
\includegraphics[width = 15cm]{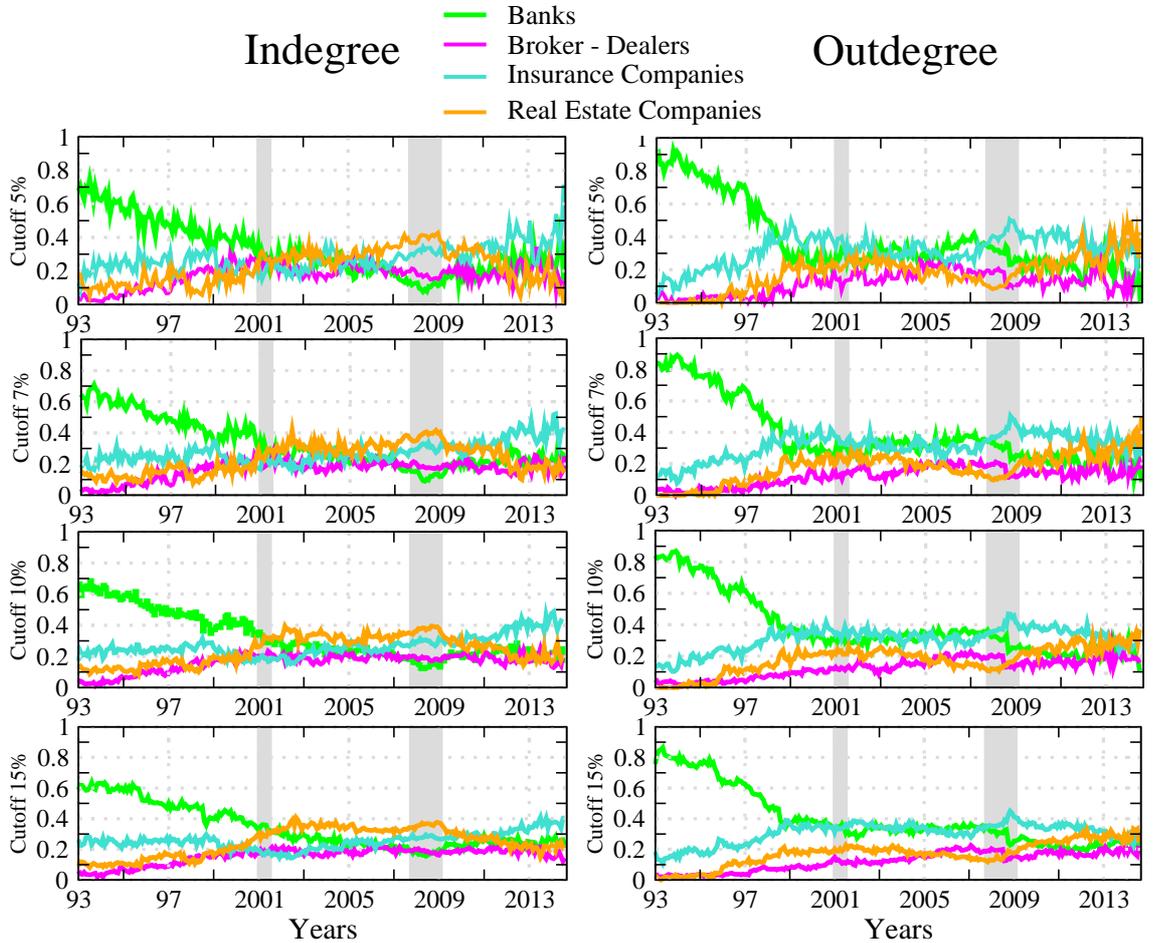}
\caption{{\bf The left (right) panel displays the relative in-degree (out-degree) per sector, from April 1993 to November 2014}: number of incoming  (outgoing) 
links  of all the nodes attached to a sector,  divided by the total number of links - banks (green), broker-dealers (purple), insurance companies (blue) and 
real estate companies (yellow). The results are reported for different values of the sensitivity parameter (i.e. test cutoff 
levels for detecting significant links). The transparent grey corresponds to US recessions as defined by the NBER.}
\label{fig:3}
\end{figure}

\noindent Next, we adopt another perspective by considering directed linkages between the set of institutions constituting our networks. Fig.~\ref{fig:3} reports
the share of incoming (in-degree) and outgoing (out-degree) connections per sector\footnote{The share reported for a sector is computed as the ratio 
between (i) the total number of incoming (outgoing) links attached to the nodes belonging to this sector, that is the in- (out-) degree associated to those
nodes, and (ii) the total number of links in the system.}. Because our directed links depict causal relationship between stock 
market returns in the sense that they differentiate the transmitter from the receiver of the financial stress, the in-degree 
allows measurement of how many institutions (from the whole sample) are affecting that sector whereas the out-degree accounts 
for how many of them are being affected by that sector. In accordance, the former measure characterises the sector 
vulnerability and the latter its influence. To ease temporal comparisons, we propose a slightly modified version of the raw in- and out-degree centrality 
measures as we normalise all the values by the total number of links in the system at each given point in time. This measure 
can be analyzed in level or in
variations to visualize more easily periods of acceleration or deceleration in network evolution. For the sake of simplicity, we keep the analysis in
level in what follows.  A visual inspection of Fig.~\ref{fig:3} shows a first notable feature for out-degrees: the banking sector experienced more pronounced changes
than other sectors, especially in the 90s. This observation is in line 
with previous discussions in the literature about the large-scale reshaping of the banking industry amid increased competition,
consolidation, and efficiency gains. As described in \cite{Calomiris}, banks embraced the new approach to client-based 
universal banking during this period, leading to a merger wave in the US banking sector. This change affected the number of 
banks in the market but also deeply affected their functioning. Our data indicates that it went along a diminution of the 
influence of the whole sector on the system.

A second notable feature is the occurrence of a shift point at the beginning of 1999. From this date
onwards, the drop in the banking sector’s share of outgoing links more or less stopped, to remain stable at around  30\% and 
40\%. By contrast, the insurance companies and, to a lesser extend, real estate firms, followed an opposite pattern. Both
sectors experienced an increase in their influence until the late 90s before stabilising around  30\% and 35\%. From early 
2000, the outgoing links of the four sectors have been kept in close ranges with a slight domination of insurance companies 
which can be deemed in this respect to be the most influential sector in the system, especially once the $2007$ - $2009$ crisis
burst out. The time series for incoming links offer a different picture. For any of the four reported figures, the cycle of 
upward and downward trends is less pronounced than for outgoing links as is the heterogeneity across sectors. Banks are still 
the most connected during the first years of the sample, exhibiting a ratio of incoming links over total links 
(relative in-degree) of about 60\%-80\% as compared to about 10\%-30\% for other sectors. However, their role as an 
important receiver of spillovers in the system kept decreasing until 2010 before slightly bouncing back.  This result 
complements the well-known dramatic fall in the number of US banks failing over the 90s as reported by the Federal Deposit 
Insurance Corporation.\footnote{See https://www.fdic.gov/bank/analytical/banking/2006jan/article2/fig5.htm.} 

Another interesting result lies in the central role played by the real estate sector which appears as the most exposed sector
between the two crises. Hence, from 2001 to 2009, its ratio of incoming links over total links was around 30\% compared to 
20\% on average for the others. This finding means that part of the underlying risk borne by real estate companies before 
the 2007 crisis was detectable by using this indicator. At this stage however, we should be careful in our conclusions as
our measurement of sector-based influence and vulnerability embeds two different effects: a size effect due to the relative
importance of the sector in the system and an individual effect corresponding to the connectedness of the nodes populating 
the sector. Part of the former results for the banking sector in Fig.~\ref{fig:1} to Fig.~\ref{fig:3}  for instance can be 
at least partly driven by the fact that many banks have disappeared over the sample.  The number of real estate companies 
and broker-dealers has grown over the years, whereas the number of insurance companies has remained constant. Accounting for 
these underlying changes in the sample could in turn change the picture. 
 
\begin{figure}[!h]
\includegraphics[width = 15cm]{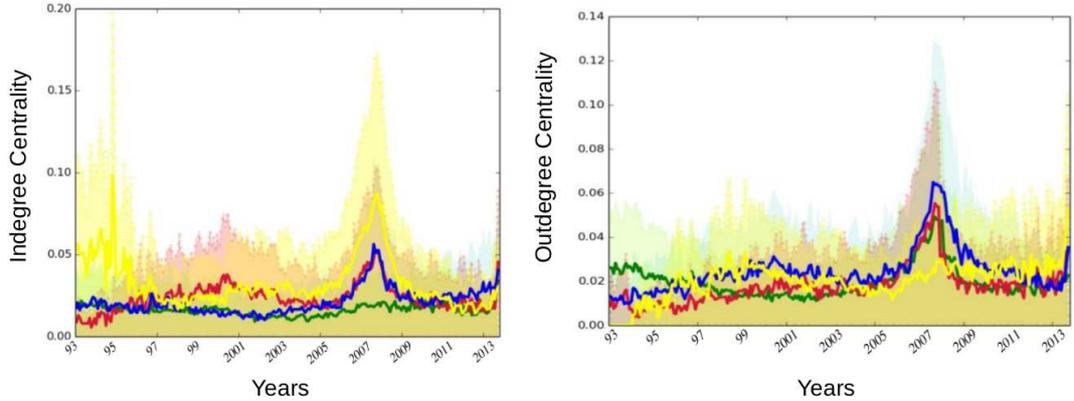}
\caption{{\bf The left (right) panel displays the  averaged In-degree (Out-degree) Centrality, from April 1993 to November 2014}: 
averaged over sectors - banks (green), broker-dealers (purple), insurance companies (blue) and 
real estate companies (yellow).  The variance of the centrality values is reported in shallow with the same colours. 
}
\label{fig:4}
\end{figure}

The next set of measurements builds on those presented in Fig.~\ref{fig:3}, while correcting for the size of each sector. It 
is computed as the average per sector of outgoing (resp. incoming) links divided by the total number of links. By doing so,
we want to abstract any potential size effect (i.e. the fact that some sectors populate the system in greater magnitude)  to 
assess the influential (resp. vulnerable) nature of the components of each sector as opposed to the actual influence (resp. 
vulnerability) of the sector as a whole. Fig \ref{fig:4} displays the evolution of the in- and out-degree 
centrality   measurements along with their variance. High values of variance are signs of broad dispersion (that is, 
heterogeneity among firms regarding that specific dimension) within each sector. We can observe that it peaks when both crises
occur. A second notable feature is that while the banking sector remains the most influential sector in the early 90s, the 
picture regarding its vulnerability with respect to the rest of the system changes markedly once we control for the size. 
Now, it does not appear as different from others. This result illustrates the interest of computing various measurements as 
they can provide different information. The way to interpret this finding is that banking institutions were not  more 
vulnerable than  other financial actors when taken individually. However, because the sector was the largest (i.e. had the 
highest number of connected actors) in the financial industry at the time, it was the greater receiver of spillovers. The 
figures regarding relative incoming links (Fig.~\ref{fig:3}) were therefore mainly driven by a size effect. Another point 
calling for attention is the bell-shape curve observed at the time of the $2007$-$2009$ financial crisis for all the series, with 
the notable exception of the banking sector relative to incoming links. Such a pattern confirms increased connectedness in the
run-up to the crisis and a fragmentation of the US financial network in the aftermath of the Lehman Brothers collapse. If we
compare the respective position of each series, our results also confirm that the banking sector was mainly a transmitter of
risk rather than a receiver during the crisis, the real estate companies appearing as the main vulnerable element of the
system with the highest average of incoming links. Eventually, the figure unveils a novel feature about insurance companies
which exhibit the highest level of influence on the rest of the system from 2000 to 2010.

\begin{figure}[!h]
\includegraphics[width = 15cm]{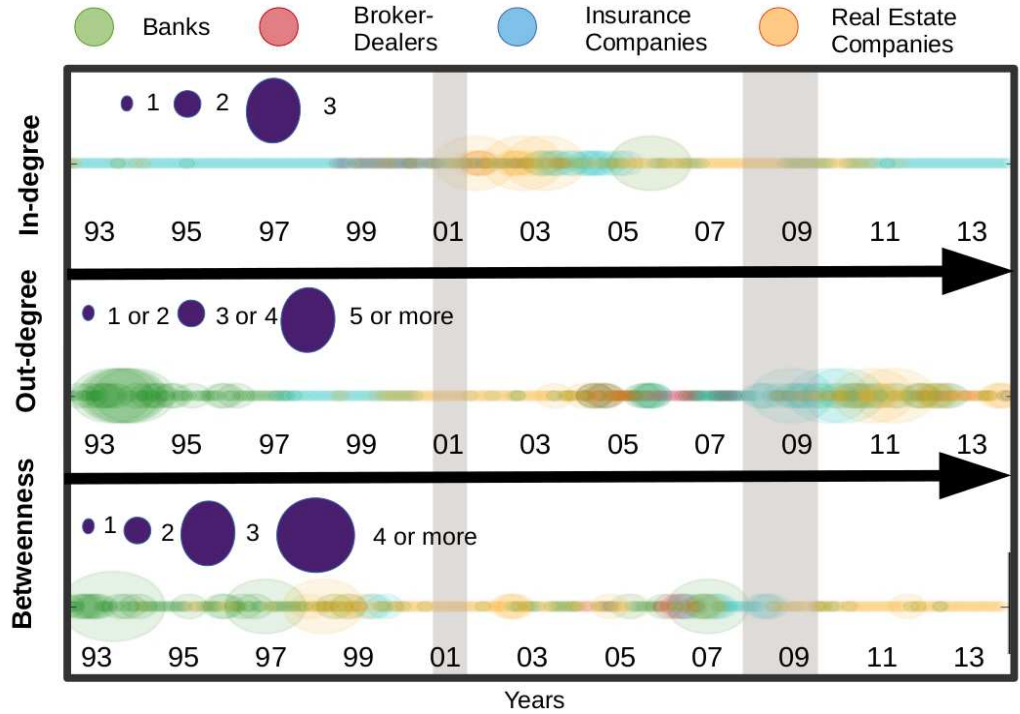}
\caption{{\bf In-degree, out-degree and betweenness top rank sectors, from April 1993 to November 2014.} The size of the circles represents the number of 
institutions with in-degree, out-degree and betweenness among the top $20\%$ highest values at each period.}
\label{fig:5} 
\end{figure}

We now focus on the most extreme cases by considering the concentration of institutions within the top 20\% values of the two
metrics discussed above. For this, we consider  for each month the different centrality measurements across institutions. We rank 
all the values in descending order. We take the first one (i.e. highest value), then, we create a threshold equal to 80\% of
the highest value. The range between the highest value and the threshold corresponds to the top 20\% highest values. Next, we
count the number of institutions per sector falling into the top 20\%. The outcome of the procedure measures   the level of 
concentration of institutions among the highest centrality values for a given month.  It also indicates whether the most 
central institution is isolated or part of a group.  If the resulting values are low, for instance, it means that one 
institution stands out above the others in term of centrality. Conversely, if they are high, it means that several 
institutions potentially belonging to the same sector are important contributors to the connectivity of the system.
When centrality is interpreted as a source of risk it means that the risk is not driven by a single institution but by a set
of institutions. Fig.~\ref{fig:5} displays the results for relative in and out-degree centrality along with the traditional 
betweenness centrality measure. For instance, it shows that among the top 20\% highest values of relative outgoing links, three 
institutions (nodes) were associated to banks in 1994 and three to real estate companies in 2011.  

Therefore, just a few institutions stand out in terms of centrality during these two periods and among 
them we counted several banks in the first years of the sample and several real estate companies at the end. More generally, 
we observed groups of influential institutions instead of isolated top institution in the early 90s and in the banking sectors
before shifting to real estate and insurance companies in the aftermath of the $2007$ - $2009$ crisis. Such a feature illustrates 
how such representation can be of interest for documenting and analysing sector-related patterns versus institution-specific 
patterns. Turning to relative incoming links, the concentration of the main receivers reached its highest level between the 
two crises with a dominance of real estate companies and, to a lower extent, insurance companies. The role of real estate 
companies and banks is in line with previous evidence from Fig.~\ref{fig:3} and Fig.~\ref{fig:4}. Fig.~\ref{fig:5} also
provides new insights, especially regarding the vulnerability of insurance companies in the years preceding the crisis. Likewise, the figure more clearly 
points out the concentration of most vulnerable institutions in the run-up to the crisis between 2001 and 2007. The third measurement added to Fig.~\ref{fig:5}
relied on betweenness centrality. Betweenness centrality, in networks representation takes high values for
nodes with central location. Being central refers for a node to the number of minimum paths between
two other nodes in the network passing through it. It is worth noting, however, that the measure is
built on undirected links, making its interpretation difficult in the context of risk propagation analysis.
Therefore, its inclusion has to be mainly viewed as a benchmark. Our results show notable discrepancies
between the patterns that emerged from in- and out-degree centrality measures and the betweenness
centrality, stressing the influence of considering the direction of the links when constructing such metrics.
We also detect similarities. In the three cases, for instance, the most pronounced changes over the period
covered correspond to well known financial events. More specifically, we uncover a concentration of
top institutions outside the crisis periods. This feature is consistent with the idea that financial risk is
building up in periods of calm - here the risk corresponds to increased connectedness - and eventually
materialised into a sudden collapse of the system before building up again. Such correspondence between
the centrality measures and financial events tends to confirm the relevance of top representations for
analysing financial networks and systemic risk.

\subsection{Components and Communities}
In the previous sections, we have analysed groups of institutions based on their sectoral classification, which correspond to 
ex-ante information. In the subsequent section, we now consider an alternative (ex-post) approach according to which 
institutions sharing strong connections as identified by specific algorithms are included in the same group as done in 
\cite{Zhou3,Song}, for instance. More specifically, we consider two different approaches. The first one relies on the identification of specific “components”, that is sets of 
nodes that are connected by means of at least one link between any pair of them. In our temporal network, several 
institutions are deemed to belong to the same component if they are connected to at least one of the nodes included in the 
component structure. Remind that two nodes are directly connected   in our context  if there exists a statistical dependence 
between the stock market returns’ time series, associated to that pair, during that month.  For this analysis, peer-interdependence 
in stock market prices is used with a cutoff (test’s cutoff levels for detecting significant links) of $10 \%$.

Given the “soft” character of this measurement, the most useful information we can retrieve from its implementation on our 
data is concerned with the composition of the small clusters that we identify not to belong to the biggest component. These 
clusters are made of institutions independent from the global system but still connected in smaller inter-dependent groups. 
For instance, in $1993$ the biggest component included about $50$ nodes. Four small components of two or three nodes each could 
be detected, leaving about $30$ nodes unconnected or too weakly connected to be considered as such to the rest.  

\begin{figure}[!h]
\includegraphics[width = 15cm]{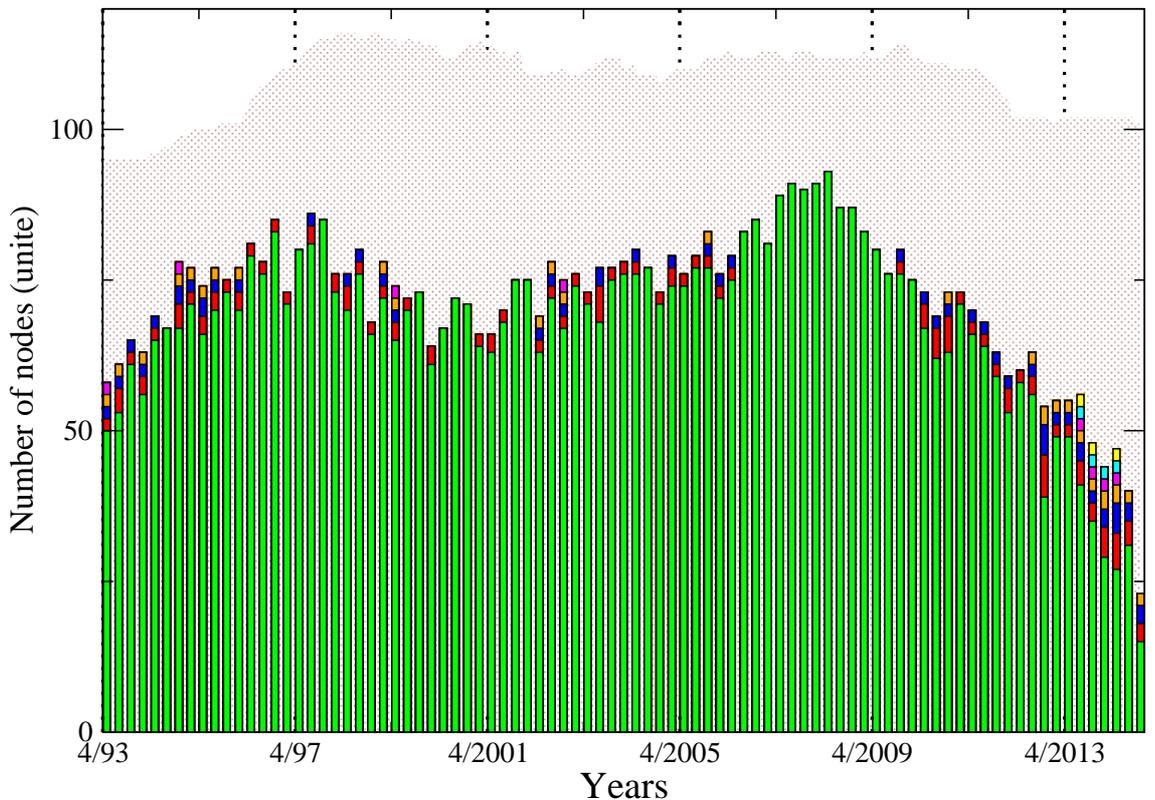}
\caption{{\bf Each bar represents the size of the components, from April 1993 to November 2014.} The dotted background depicts the 
total number of institutions for each month. The green colour is attached to the biggest component, the red colour to the 
second biggest and so on.}
\label{fig:6}
\end{figure}

\begin{figure}[!h]
\includegraphics[width = 14cm]{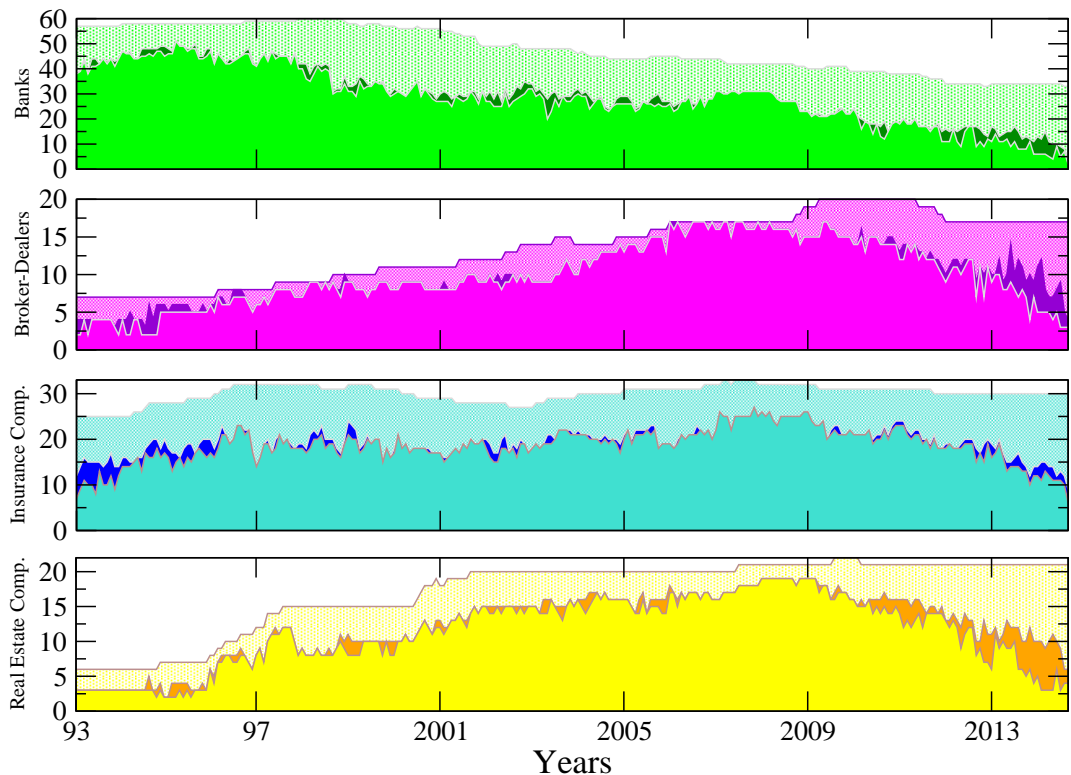}
\caption{{\bf Number of institutions per sector broken down into the number of institutions in the biggest component 
(plain-light colour), the number of institutions in small components (plain-dark colour) and the number of isolated 
institutions (dotted-light colour), from April 1993 to November 2014.}}   
\label{fig:7}
\end{figure}

\begin{figure}[!h]
\includegraphics[width = 14cm]{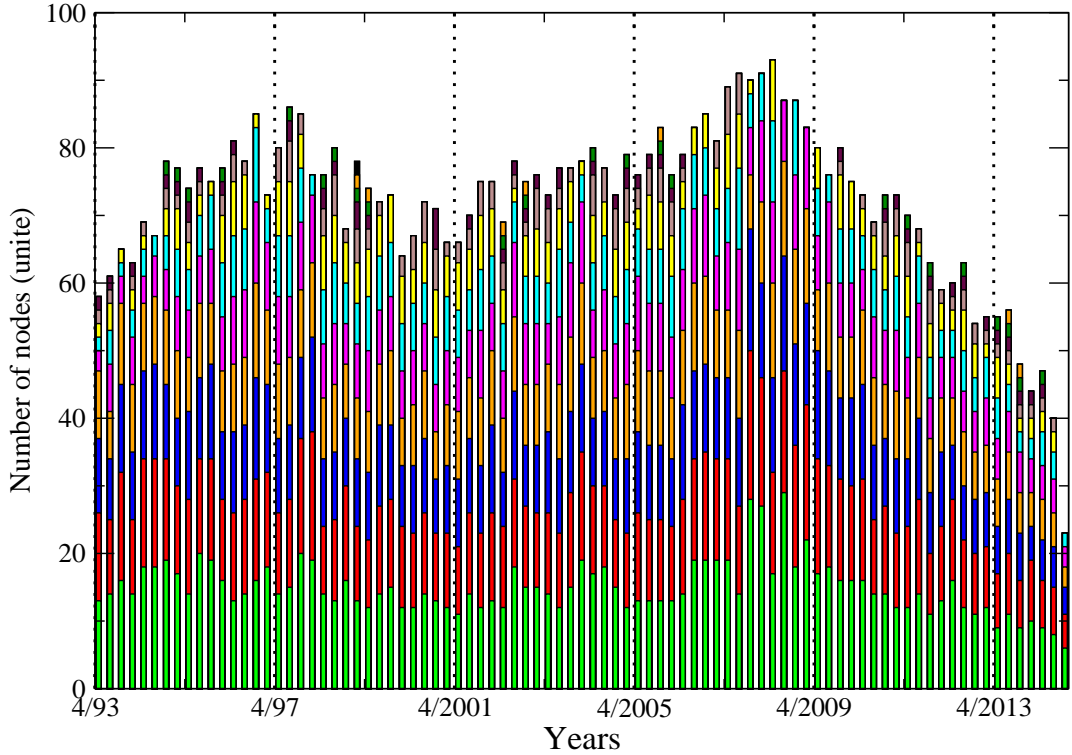}
\caption{{\bf Each bar represents the number of institutions making up each community within the giant component, from April 1993 to November 2014.} From 
bottom to top, the green colour is attached to the biggest community, the red colour to the second biggest and so on.}
\label{fig:8}
\end{figure}

The first notable observation in  Fig.~\ref{fig:6} is the agglomeration of almost all the institutions in the biggest
component over the whole sample, denoting global dependence of institutions among financial companies in the US. Yet, the 
number of non-connected  or isolated, institutions is not negligible, as represented in dotted background. In most of  cases,
there are also a few institutions that are gathered in very small groups. The $2007$-$2009$ crisis period is particularly 
interesting. We observe a growing integration of the US industry as materialised by the detection of only one single large 
component in $2007$ and $2008$ as opposed to two, three or four small components in the years before. During this period, 
the largest component attains its sample peak at $90$ (for a total of about 110 nodes at this date). Since then, a 
fragmentation process has been ongoing until $2014$ with both more components and an abrupt and monotonous drop in the total 
of connected nodes.
\\
Fig.~\ref{fig:7} provides complementary insight by combining pieces of information on the components and on the sectors. Two
features emerge from this exercise. First, the picture provided by the banking sector is contrasted. On the one hand, most of 
the institutions from that sector are part of the biggest component and only very few of them belong to smaller components. On
the other hand, many banks and insurance companies appear isolated. The situation is more balanced for other sectors. 
Broker-dealers and real estate companies are mainly integrated within the biggest component at the beginning of the sample 
while a non-negligible part of these institutions move to small components or become isolated in the post-crisis period. A 
detailed study about the formation (and stability) of the components is out of the scope of the present study, but this 
analysis might be of interest for future research.
\\
The classification in components is related to the occurrence of at least one link between pairs of nodes. A more nuanced way 
to gather the nodes into relevant groups of interacting entities relies on the identification of community structure. A 
community is a set of nodes more connected between them than with the rest of the system. There exist several algorithms to 
identify community structures. Our community detection was performed relying on the Louvain algorithm. \cite{Blondel2008} 
\footnote{http://perso.crans.org/aynaud/communities/}. 
The goodness of a partition is measured by the modularity. For the undirected version of our networks (where the community 
detection has been performed), optimising modularity can be interpreted both, as optimising a particular stochastic block
model and a particular diffusion process on the networks  \cite{Schaub}. The best partition is the one that maximises 
modularity. In our causality-based networks, institutions grouped into communities means that their market returns display
strong temporal dependences. Fig. 8 reports the number of institutions constituting communities within the biggest component.
We recall that a detailed analysis of the dynamic nature of those communities and their determinants over time is out of the 
scope of the study. However, we can make some general remarks from this figure, such as the presence of a higher number of 
communities than the number of sectors, showing that both pieces of information are not redundant. Also, it seems that the
community structures exhibit strong stability over time. For instance, we do not observe the same downward slope pattern that
was noticed for the number of connected banks in Fig.~\ref{fig:8}. We do observe, nevertheless, a slight peak around 
the time of the 2007-2009 crisis, characterised by a small  decrease in the number of communities and an increase in the number of institutions included 
in the biggest components - sum of institutions across communities,  meaning that communities size expanded during this period. Interestingly, the patterns observable in Fig.~\ref{fig:8} are very
much comparable to those discussed for Fig.~\ref{fig:6}, that is, higher integration during the 2007-2009 crisis and increased fragmentation afterwards, 
characterised by the emergence of many clusters of institutions. 
\subsection{Interface}
\begin{figure}[!h]
\begin{center}
\includegraphics[width = 12cm]{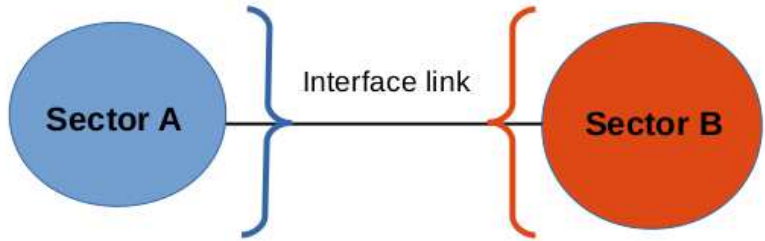}
\end{center}
\caption{{\bf A link is defined as part of the interface if its two connecting nodes belong to different sectors.}}
\label{fig:9}
\end{figure}

\begin{figure}[!h]
\includegraphics[width = 14cm]{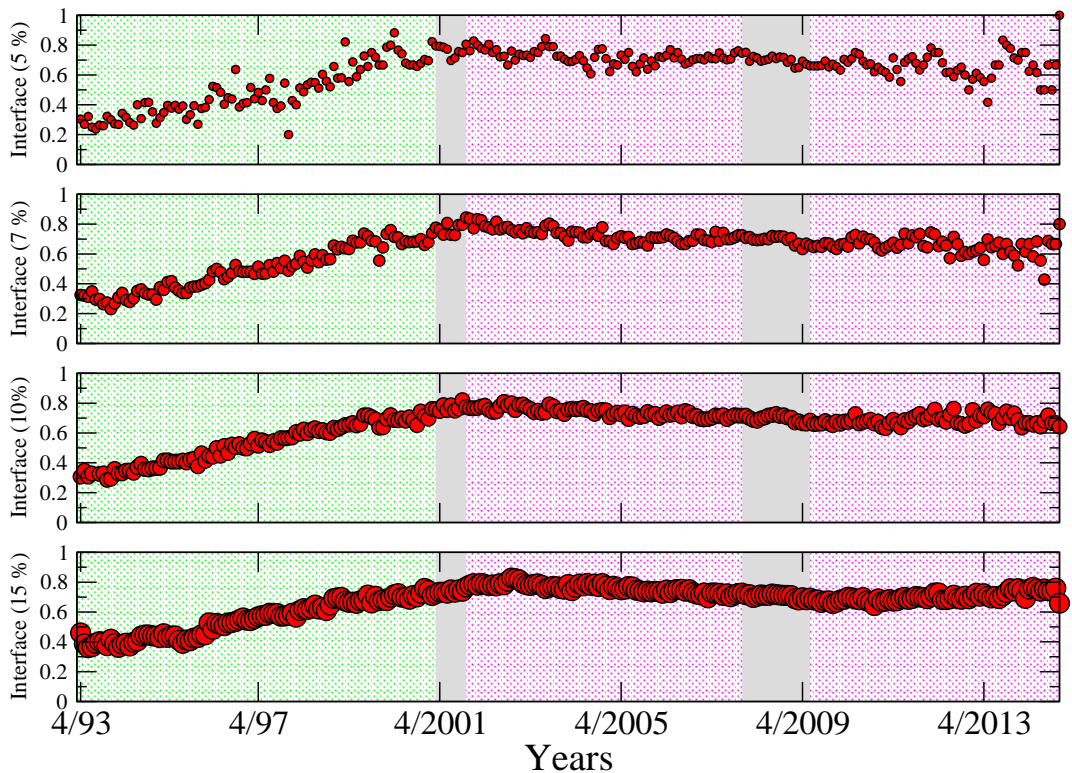}
\caption{{\bf Proportion of sector-interface inside the giant component from April 1993 to November 2014, for different values of the test\textquoteright s cutoff 
levels for detecting the significant links.} Two different 
background colours are used to indicate two different phases: in green, a monotonous increasing phase, and in pink, a stable
phase around 0.8. It is important to notice that the rise is not a consequence of increments in the number of links, as the
values showed are normalised. The measurements are shown for different values of the sensitivity (i.e. cutoff) parameter.}
\label{fig:10}
\end{figure}

This section aims to complement previous analysis by combining information on institutions' linkages and their sector-based
classification. To that end, we use the notion of \textquotedblleft sectoral interface\textquotedblright. We define a link as
being part of the interface if it connects two nodes belonging to different sectors (see Fig.~\ref{fig:9}). In Fig. 
\ref{fig:10} we show the proportion of sectoral interface inside the largest component. Two different background colours are 
used to indicate two successive phases. In green is a monotonous increasing phase, where the proportion of links between 
different sectors is growing until the first 2001 crisis, starting from around 35\% in 1993 to 80\%. After this turning point,
in pink, the proportion of inter-sectoral links stabilise while keeping fluctuating around $80\%$. It is important to note 
that the rising pattern is not a symptom of an inflated number of links in the system, as the values shown are normalised. 
The measurements are shown for different values of the sensibility parameter ($5,7,10 $ and $15 \% $), in order to illustrate 
the robustness of the results. The pattern that we identify here in sectoral interface illustrates the high level of 
integration of the financial industry and in particular, the presence of strong connections across sectors that can be 
traced back long before the occurrence of the 2007-2009 crisis. Also, it shows that the level of integration across sectors
was far lower in the 90s, a decade which experienced several large-scale bankruptcy episodes in the financial sector such as 
the failure of LTCM with less comparable impact on the real economy than the 2000s crises. 
The transition phase that we have been able to identify echoes previous results in the literature reporting the existence of 
similar transition phenomena in economics \cite{Saracco,Stanley2014}.
\subsection{M-reach centrality (contagion processes) vs. Katz centralities}
We complete our analysis of  the financial industry by going beyond direct connections and considering indirect ones.  To 
this end, we use the   Katz centrality  measure as in \cite{Cohen}, along with the contagion process which has been more 
considered in financial application. Katz centralities are measurements used to understand the importance of each node. It 
measures the connections of a node with other nodes of the network through both direct and indirect links. The weight attached 
to each direct and indirect connected nodes is a function of the distance between the two nodes. In accordance, the more 
indirect the connection - that is, the more intermediate nodes - the higher the distance and the lower the weight. Formally, 
the distance is featured through a penalising attenuation factor. The Katz centrality is a good alternative to eigenvector 
centrality for measuring centrality beyond first-degree nodes when the network is sparse or directed.

\begin{figure}[!h]
\centering
\includegraphics[width = 13cm]{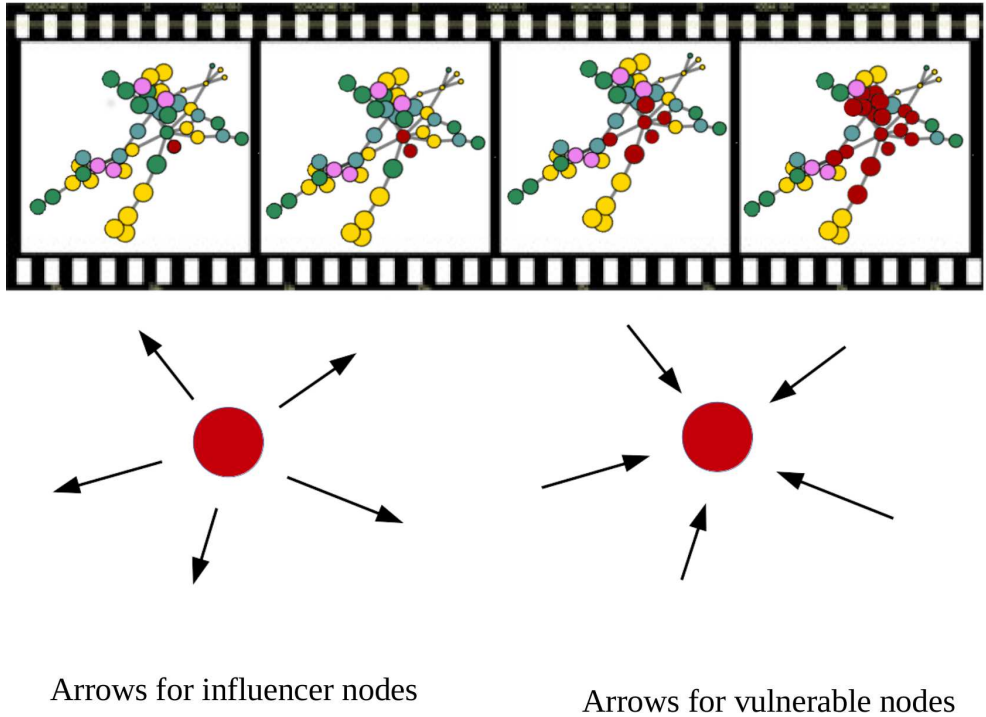}
\caption{{\bf Infection process}. Top figure: Infection process in three steps to calculate the biggest influencer and vulnerable nodes. 
Bottom figure: Direction of the arrows to calculate the influencer institutions (left) and the vulnerable nodes (right).}
\label{fig:11}
\end{figure}

An alternative and  intuitive approach to Katz centrality is achieved by applying percolation theory and computing the m-reach
centrality. In the context of outgoing links, it builds on a very similar principal to the Katz centrality measure by 
identifying how   influential each node is in its ability to spread infection or shocks -  in our case severe losses -  to 
the rest of the system. In our simple contagion process, we start with an infected node as an initial seed and then we count 
the number of nodes touched by this infection through outgoing links (see top panel of Fig.~\ref{fig:11}). We can illustrate
the algorithm with the following example. Starting from one node with three outgoing links, three nodes are infected in the 
first step. In the second step, the three infected nodes become diffusers of the disease to their neighbours via their 
outgoing connections and so on. We stop the procedure at the third iteration and count the total number of infected nodes. 
Note that the contagion process will highlight the structural role of nodes. For instance, the target node in 
Fig.~\ref{fig:11} is weakly connected to the network and with a peripheral role; however, its connectivity to a highly central
hub contributes structurally to global connectivity. To detect vulnerable nodes, we apply the same procedure, having initially 
inverted the direction of all the arrows in the networks (see bottom panel of Fig.~\ref{fig:11}). After three steps, the
higher the number of infected nodes, the higher the number of institutions influencing the stock market price of the 
institution from which we started. The effect of inverting the arrows over the contagion process gives the number of 
influencer institutions each node is vulnerable to, then characterising its level of fragility. 

\begin{figure}[!h]
\includegraphics[width = 15cm]{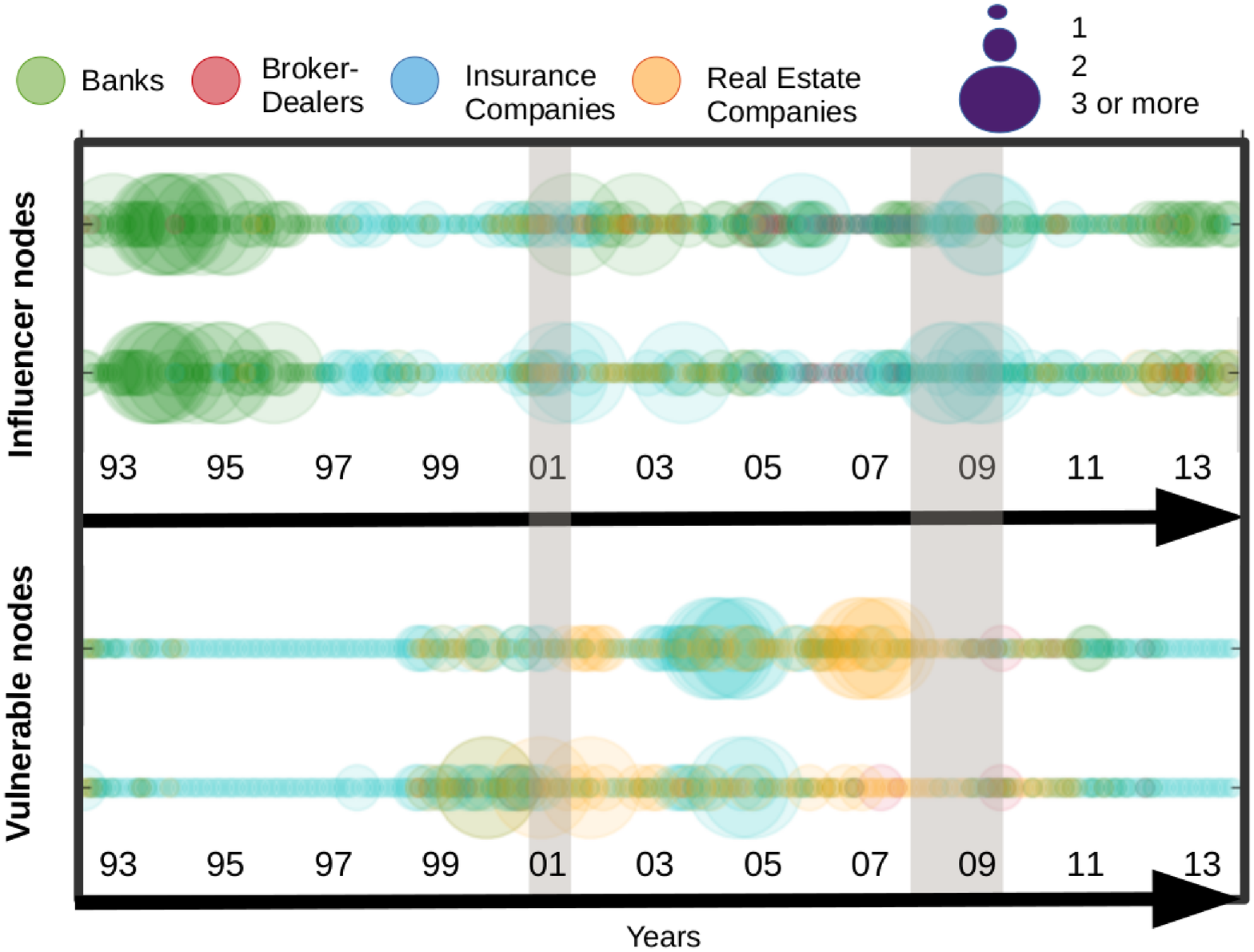}
\caption{{\bf Most influencer and vulnerable nodes by means of a contagion process in the upper part, followed by the Katz 
centrality, from April 1993 to November 2014.} The size of the circles represents the number of institutions with out-degree and in-degree among the top 20\% highest values of the sample.}
\label{fig:12}
\end{figure}

 In Fig. \ref{fig:12}, we compare the results obtained by means of our contagion process  with the Katz centrality measure. We 
 stop the contagion process at the second iteration. For both measures, we count the number of institutions per sector which 
 fall into the top 20\% highest values.  As explained in a previous section, the top 20\% embeds the values ranging from 
 the highest centrality value  in a given month to a threshold which is equal to 80\% of the highest value.  This  procedure 
 is repeated at every period to flag the most influential/vulnerable institutions over time.  The circles in the top panel, 
 display the influence of each sector  for the contagion process.  Those for the Katz   centrality stand just below.   A
 visual inspection of the two subfigures support the closeness of their information content as the patterns are very similar
 over the whole sample. We can nevertheless notice a few differences.  Among them, we can cite the role of the banking sector
 before the 2001 crisis which is characterised as a substantial propagator of spillovers only when using the Katz centrality 
 measure. Turning to nodes’ influence, we can see that both results again provide similar information: (i) the banking sector 
 is highly influential from 1993 to 2001, (ii) the insurance companies along with banks have emerged over time as key
 transmitters in the financial industry.
 
 \subsection{Temporal measurements}
\begin{figure}[!h]
\includegraphics[width = 15cm]{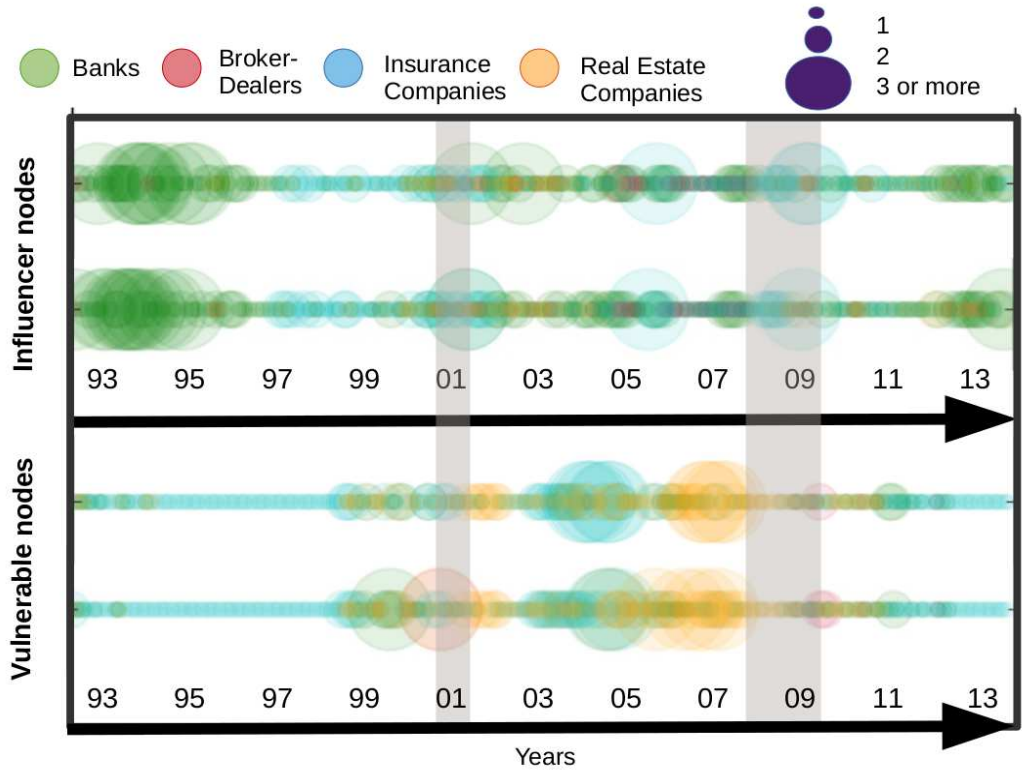}
\caption{{\bf  Most influencer and vulnerable nodes by means of an instantaneous contagion process (static 2-reach centrality 
measure), in the upper part, and a contagion process with time delay (temporal 2-reach centrality measure),  in the 
lower part from April 1993 to November 2014.} The size of the circles represents the number of institutions with out-degree 
and in-degree among the top 20\% highest values of the sample.}
\label{fig:13}
\end{figure}

 More recently, a new line of research in network science has aimed to develop specific metrics for dynamic systems as opposed
 to standard metrics applied on successive static snapshots of time evolving networks. In accordance, we propose  to apply as
 a last exercise a metric embedding information on the temporal sequence of edges and nodes. Fig. \ref{fig:13} displays both 
 temporal and a static metric. The former is computed by sector as the sum of top 20\% most central institutions. We do it for
 both in-degree (i.e. \textquotedblleft vulnerable\textquotedblright\ nodes) and out-degree (i.e. \textquotedblleft 
 influencer\textquotedblright\ nodes).  We detail below how out-degrees and in-degrees are computed to account for time 
 variation in the connections. Starting with in-degree, at each point in time, we first count for each node the number of 
 incoming links at time $t$. We then count the number of incoming links at the previous period, i.e. at $t-1$, of the 
 connected nodes. We can view this measure as a modified version of the M-reach centrality in which  first order connections
 and second order connections stem from networks at two successive time periods. By doing so, the metric features potential 
 time delay in the contagion process and account for the sequence of appearance and disappearance of links over time. The 
 reverse logic is applied for out-degree. At time $t$, we count the number of outgoing links (i.e. connected nodes). By doing 
 so, we identify for each node a set of influenced nodes (i.e. nodes from the network connected through outgoing links). Then,
 at $t+1$, we count the number of links of the connected nodes. The spirit of such a measure is to the one of an infection
 process with time delay in which for instance first order neighbors are infected  at time $t$ considering the state of the 
 network at time $t$ and then second order neighbors are infected at time $t+1$ considering the state of the network at time 
 $t+1$. Fig. \ref{fig:13} also displays, above each temporal metric, the previous static version of the M-reach centrality
 metrics that is when we count the number of second order neighbors at fixed time period.  We observe that both temporal and 
 static metrics provide consistant information regarding the vulnerability and influence of sectors. We also note a
 few differences. A notable divergence for instance lies in the high concentration of vulnerable sectors in the run up to 
 the 2007-2009 financial crisis that appears with the temporal metric and is less visible with the static one.

\section{Conclusion and Discussion}
Using a large set of tools from networks science, causality-based networks have been analysed in a large set of $155$ financial 
institutions: all the banks, broker-dealers, insurance and real estate companies listed in the Standard \& Poor’s $500$ index 
during the period $1993$ - $2014$. In contrast to main body of research on financial networks, we pay particular attention to its 
temporal dimension by following the approached used in \cite{Geraci} designed to deal with its dynamic nature of financial institutions’
connections. Equipped with these dynamic causality-based networks, we describe its evolution per sector in the vein of \cite{Billio2} by 
using traditional tools from networks science such as centrality measures based on in-degree and out-degree as well as more
advanced tools. The latter are intended to expend further traditional analysis by extracting information that was not 
attainable using simple sectoral centrality measures. Among the set of tools we are using, we can specifically emphasise 
measures of community and component structures as well as interface identification to offer a different view on the 
fragmentation/integration processes that took place over time in the US financial industry. We also apply an algorithm 
derived from the percolation theory to shed light on the question of influencing/vulnerable nodes or groups of nodes. 
Eventually, we propose a top institution representation drawing on the most highly connected institutions.  By doing so, we 
can provide original empirical insight and tackle the following two objectives of the paper: (i) to provide new insights to 
network science by the mean of financial data, and (ii) to improve our understanding of the US financial industry over a long 
time span. Regarding our first objective, our work is one more attempt to construct a bridge between the physics thinking, in
the spirit of simplicity, and equivalence between measurements. From the comparison of the different measures, our results 
tend to show that a large set of information can be extracted from the traditional in and out-degree centrality measures at 
three different levels: as (i) node-level, (ii) sectoral level, and (iii) by considering top values. Further information can
be extracted from financial data by means of the communities’ and components’ structures. Eventually, our results suggest 
the quasi-equivalence on financial data of alternative measures as the one built on contagion process and the Katz centrality.

Turning to the second objective related to the systemic risk analysis, our results allow us to document four important 
patterns. First, banks have been highly influential since the early 1990s as documented by the temporal evolution of our 
normalized out-degree measurement as well as the contagion process. Second, real estate companies have been the most 
vulnerable sector in the financial industry especially during the period $2001$ - $2007$. This is illustrated by in-degree 
centrality measurement and the (inverted) contagious process. Third, market integration drastically increased in the run-up to
the $2007$ - $2009$ crisis either within the financial sector or between them. Component as well as community structures provide
clear evidence on this feature. Fourth, the US financial industry has experienced a growing fragmentation from the crisis to
the late $2013$. This pattern appears clearly from the analysis of component and community structures.  Finally, we applied a temporal measure, where the previously developed contagious process is applied but now over consecutive
times windows. This last metric can be considered to study cascading failure mechanism. 

Such results open up two important observations. First, it confirms in a dynamic context that various tools from networks 
science can improve our knowledge of the financial system, stressing the need for further research in this direction. Second,
the identification of a high and persistent level of integration across sectors calls into question the current sector-based 
approach to macroprudential surveillance.

 This research can be extended in different ways. First, we rely in the whole study on an unweighted network as done in \cite{Billio2}. A natural extension would be to explore whether additional information is embedded in the intensity of links. Second, in line with a recent strand of the literature in network science and as illustrated in the last subsection of the analysis, using specific metrics dedicated to temporal network appears as a promising line of research to analyse financial systems.


\section*{Supporting Information: Appendix SA. Expansion of section Methods: Considerations about the prior distribution and the posterior distribution 
sampling algorithm model.} 

\section*{Acknowledgments}
Computational resources have been provided by the Consortium des \' equipements de Calcul Intensif (CBI), funded by the Fonds 
de la Recherche Scientifique de Belgique (F.R.S.-FNRS) under Grant No. 2.5020.11. We also gratefully acknowledge financial 
support from the Communaut\' ee fran\c caise de Belgique under the ARC Grant No. 13/17-055. We thank Renaud Lambiotte and 
Jean-Charles Delvenne as well as the participants to the BeNet Conference in 2016 at Louvain la Neuve as well as the participants to the 
research seminar at Ecole Polytechnique (UCL) for their feedbacks. YG thanks to Mauro Faccin and Timoteo Carletti for valuable discussions.


\begin{thebibliography}{100}
\bibitem{Geraci}
Geraci M, Gnabo JY (2017) Measuring interconnectedness between nancial
  institutions with bayesian time-varying vector autoregressions.
\newblock Journal of financial and quantitative analysis (forthcoming) .

\bibitem{Basel}
on~Banking~Supervision BC (2013) Global systsystemic important banks: updated
  assessment methodology and the higher loss absorbency requirement.
\newblock Tech rep, Bank for International Settlements .

\bibitem{bouchaud2008}
Bouchaud JP (2008) Economics needs a scientific revolution.
\newblock Nature 455: 1181-7.

\bibitem{farmer2009}
Farmer D, Foley D (2009) The economy needs agent-based modelling.
\newblock Nature 460: 685-686.

\bibitem{Billio}
Billio M, Getmansky M, Lo AW, Pelizzon L (2012) Econometric measures of
  connectedness and systemic risk in the finance and insurance sectors.
\newblock Journal of Financial Economics 104: 535-559.

\bibitem{Diebold}
Diebold FX, Yilmaz K (2014) On the network topology of variance decompositions:
  Measuring the connectedness of financial firms.
\newblock Journal of Econometrics 182: 119-134.

\bibitem{Zhou3}
Dai YH, Xie WJ, Jiang ZQ, G-JJiang, Zhou WX, et~al. (2016) Correlation
  structure and principal components in the global crude oil market.
\newblock Empirical Economic 51: 1501-1519.

\bibitem{Zhou4}
Han RQ, Xie WJ, Xiong X, Zhang W, Zhou WX (2017) Market correlation structure
  changes around the great crash: a random matrix theory analysis of the
  chinese stock market.
\newblock Fluctuation and Noise Letters 16: 1750018.

\bibitem{Primiceri}
Primiceri G (2005) Time varying structural vector autoregressions and monetary
  policy.
\newblock Review of Economic Studies, 72: 821-852.

\bibitem{Billio2}
Bianchi D, Billio M, Casarin R, Guidolin M (2015) Modeling contagion and
  systemic risk.
\newblock SSRN Working paper .

\bibitem{Gandica}
Gandica Y, Lambiotte R, Carletti T (2016) What can wikipedia tell us about the
  global or local character of burstiness?
\newblock The Workshops of the Tenth International AAAI Conference on Web and
  Social Media Wiki: Technical Report WS-16-17 .

\bibitem{Cogley}
Cogley T, Sargent T (2005) Drifts and volatilities: Monetary policies and
  outcomes in the post wwii us.
\newblock Review of Economic Dynamics 8: 262-302.

\bibitem{Calomiris}
Calomiris CW, Karceski J (1998) Is the Bank Merger Wave of the 1990s Efficient?
  Lessons from Nine Case Studies.
\newblock AEI Press.

\bibitem{Koop}
Koop G, Leon-Gonzalez R, Strachan RW (2010) Dynamic probabilities of
  restrictions in state space models: An application to the phillips curve.
\newblock Journal of Business and Economic Statistics 28: 370-379.

\bibitem{Song}
Song DM, MTumminello, Zhou WX, Mantenga R (2011) Evolution of worldwide stock
  markets, correlation structure, and correlation based graphs.
\newblock Phys Rev E 84: 026108.

\bibitem{Blondel2008}
Blondel V, Guillaume JL, Lambiotte R, Lefebvre E (2008) Fast unfolding of
  communities in large networks.
\newblock Journal of Statistical Mechanics 10: P10008.

\bibitem{Schaub}
Schaub MT, Delvenne J, Rosval M, Lambiotte R (2017) The many facets of
  community detection in complex networks.
\newblock Applied Network Science : 4.

\bibitem{Saracco}
Saracco F, Clemente RD, Gabrielli A, Squartini T (2016) Detecting early signs
  of the 2007–2008 crisis in the world trade.
\newblock Journal Scientific Reports 6: P30286.

\bibitem{Stanley2014}
Meng H, Xie WJ, Jiang ZQ, Podobnik B, Zhou WX, et~al. (2014) Systemic risk and
  spatiotemporal dynamics of the us housing market.
\newblock Scientific Reports 4: 3655.

\bibitem{Cohen}
Cohen-Cole E, Patacchini E, Zenou Y (2011) Systemic risk and network formation
  in the interbank market.
\newblock CEPR Discussion Paper No DP8332 .

\end{thebibliography}

\end{document}